\documentclass[11pt,preprint,superscriptaddress,aps,showkeys,nofootinbib]{revtex4}
\usepackage{amssymb,amsmath,amsfonts}
\usepackage{graphicx}
\usepackage{graphics}
\usepackage{eepic,epsfig}
\usepackage{indentfirst}
\usepackage[utf8]{inputenc}
\usepackage[T1]{fontenc}

1.60cm \makeatletter \@addtoreset{equation}{section}

\makeatother

\begin{document}
\title{Casimir effects in Lorentz-violating scalar field theory}
\author{M. B. Cruz}
\affiliation{Departamento de F\'{\i}sica, Universidade Federal da Para\'{\i}ba\\
 Caixa Postal 5008, 58051-970, Jo\~ao Pessoa, Para\'{\i}ba, Brazil}
\email{messiasdebritocruz@gmail.com, emello, petrov@fisica.ufpb.br}
\author{E. R. Bezerra de Mello}
\affiliation{Departamento de F\'{\i}sica, Universidade Federal da Para\'{\i}ba\\
 Caixa Postal 5008, 58051-970, Jo\~ao Pessoa, Para\'{\i}ba, Brazil}
\email{messiasdebritocruz@gmail.com,  emello, petrov@fisica.ufpb.br}
\author{A. Yu. Petrov}
\affiliation{Departamento de F\'{\i}sica, Universidade Federal da Para\'{\i}ba\\
 Caixa Postal 5008, 58051-970, Jo\~ao Pessoa, Para\'{\i}ba, Brazil}
\email{messiasdebritocruz@gmail.com,  emello, petrov@fisica.ufpb.br}

\begin{abstract}
In this paper we consider a Lorentz-breaking extension of the theory for a real massive scalar quantum field in the region 
between two large parallel plates, with our manner to break the Lorentz symmetry is CPT-even, aether-like. For this system we 
calculated the Casimir energy considering different boundary conditions. It turns out to be that the Casimir energy strongly 
depends on the direction of the constant vector implementing the Lorentz symmetry breaking, as well as on the boundary 
conditions.
\end{abstract}

\maketitle


\newpage
\section{Introduction}

The Casimir effect, discovered by H. B Casimir in 1948 \cite{Casimir:1948dh} and experimentally confirmed ten years later by 
M. J. Sparnnaay \cite{Sparnaay:1958wg}, is one of the most direct manifestations of the existence of vacuum quantum 
fluctuations.\footnote{In the 90s, experiments have confirmed the Casimir effect with high degree of accuracy \cite{Lamoreaux:1996wh}, \cite{Mohideen:1998}.}
The interest to it strongly increased in 1970s within the context of the Quantum Field Theory (QFT).

By its definition, the Casimir effect occurs within the interaction between two uncharged conductor plates placed in the quantum vacuum.
So we can conclude that the Casimir effect is a purely quantum phenomenon. Indeed, within the classical 
electrodynamics, the interaction between two uncharged conductor plates is always zero.

In general, we can define the Casimir effect as being a stress (force per unit area) when boundary conditions are imposed 
on quantum fields. These boundaries can be material means, interfaces between two phases of the vacuum, or even, space-time 
topologies.

The simplest way in which it is possible to study the Casimir effect is the case of the interaction between two parallel plates placed in the vacuum. 
However, vacuum is an infinite set of waves which contemplate all wavelength possibilities, and when the plates are 
considered in this vacuum, only a specific wavelengths are allowed between them.

When the vacuum energy is calculated between the parallel plates, an infinite amount (ultraviolet divergences) is obtained. 
So we use the Abel-Plana formula \cite{{Bordag:2009zzd,Saharian:2007ph}} to regularize the vacuum energy. With this, the infinite energy due to the vacuum 
considering the two parallel plates is subtracted from the infinite energy of the free vacuum, resulting in a finite energy.

In the usual QFT, the Lorentz symmetry is preserved. However, other theories 
propose models where the symmetry of Lorentz is violated \cite{{Ulion:2015kjx,Ferrari:2010dj}}. Thus, the space-time anisotropy in a given QFT model should 
certainly modify the Hamiltonian operator spectrum.

In recent years the Lorentz symmetry breaking has been questioned, both in the theoretical and experimental context. In 
1989, V. A. Kostelecky and S. Samuel \cite{Kostelecky:1988zi} described a mechanism in string theory that allows the violation of Lorentz symmetry at the Planck energy scale. 
This mechanism is based on a spontaneous violation of the Lorentz symmetry, implemented
through the emergence of expected values of nonzero vacuum by some vector and tensor 
components, which implies in preferential directions, therefore, space-time anisotropy.

If there is a violation of the Lorentz symmetry at the Planck energy scale in a more fundamental theory, the effects of this 
breakdown must manifest itself in other energy scales in different QFT models. Other mechanisms of violation of Lorentz 
symmetry are possible, such as space-time non-commutativity \cite{Carroll:2001ws, Anisimov:2001zc, Carlson:2001sw,
Hewett:2000zp, Bertolami:2003nm}, variation of coupling constants \cite{Kostelecky:2002ca, Anchordoqui:2003ij, 
Bertolami:1997iy}  and modifications 
of quantum gravity \cite{Alfaro:1999wd, Alfaro:2001rb}. 

The Lorentz symmetry breaking acquired a great experimental interest. Modern experiments have shown the high accuracy 
of the results obtained through QFT. Of course, the Casimir effect credits a great phenomenon to study the violation of the 
Lorentz symmetry in theoretical models of Field Theory, therefore, in order to direct the experiments in search of vestiges 
left by the break of the Lorentz symmetry.

A great number of studies of different consequences of the Lorentz symmetry breaking on the tree level is presented in papers \cite{class}. In this context, study of the implications of the Lorentz symmetry breaking within the Casimir effect becomes very natural.
The first studies of the Casimir effect in the Lorentz-breaking theories have been carried out in \cite{FrankTuran,Kharlanov,Escobar} for different Lorentz-breaking extensions of the QED. However, up to now, there was no detailed studies of the Casimir effect specially devoted to the Lorentz-breaking extensions of the scalar field theory. In this paper, we carry out this study. The importance of studies of the Casimir effect associated with the scalar field theory is confirmed by the fact that, in the Lorentz-invariant case, it has been discussed in great details in \cite{Bordag:2009zzd,Milton}. Therefore, generalization of these studies to the Lorentz symmetry-breaking case is a very natural extension, and this analysis deserves to be developed.

In this work, we aim to provide additional theoretical predictions about the quantum vacuum in a modified Klein-Gordon model, 
in particular, we focused on calculating Casimir energy per unit area.

Our paper is organized as follows. The Section II briefly describes the theoretical model for the real scalar field considering 
an aether-like CPT-even Lorentz symmetry breaking, through the direct coupling between the derivative of the field with an 
arbitrary constant four-vector. In the Section III we focus on calculating the Casimir energies, in cases where the Lorentz 
violating vector is parallel and orthogonal to the plates, and the following boundary conditions were considered: Dirichlet, 
Neumann and mixed ones, respectively. In the Section IV we summarize the results obtained in the paper. Here, we assume 
$\hbar=c=1$, the metric signature will be taken as $(+,-,-,-)$.


\section{The Model}\label{themodel}

In this section we will introduce the theoretical model that we want to investigate. This model is composed by a massive real  scalar quantum field whose dynamics is given by Lagrangian density below:\footnote{Originally it has been introduced as an ingredient of the Lorentz-violating extension of the standard model \cite{ColKost}.}
\begin{eqnarray}
\label{lagrangedensity}
{\cal{L}}=\frac{1}{2}\left[\partial_\mu\phi\partial^\mu\phi+{\lambda} (u\cdot\partial\phi)^2+m^2\phi^2\right] \ .
\end{eqnarray} 
Here the dimensionless parameter $\lambda$ is supposed to be much smaller that one. It codifies the Lorentz symmetry violation 
of the system caused by the presence of a coupling between the derivative of the scalar field with a constant 
vector $u^{\mu}$.\footnote{At the quantum level, this model for the scalar field was considered in \cite{Petrov.10}.} 

The modified Klein-Gordon equation reads:
\begin{eqnarray}
\label{MKG1}
\left[\Box+\lambda(u\cdot\partial)^2+m^2\right]\phi(x)=0 \ . 
\end{eqnarray}

The energy-momentum tensor, as usual, is defined as:
\begin{eqnarray}
T^{\mu\nu}&=&\frac{\partial{\cal{L}}}{\partial(\partial_\mu\phi)}(\partial^\nu\phi) -\eta^{\mu\nu}{\cal{L}} \ . 
\end{eqnarray}
So, we have
\begin{eqnarray}
\label{energy-tensor}
T^{\mu\nu}&=&(\partial^\mu\phi)(\partial^\nu\phi)+\lambda u^\mu(\partial^\nu\phi) (u\cdot\partial\phi)-\eta^{\mu\nu}{\cal{L}} \  ,
\end{eqnarray}
where $\eta^{\mu\nu}$ denotes the usual Minkowski flat space-time metric tensor. We can verify that 
\begin{eqnarray}
\partial_\mu T^{\mu\nu}=0 \  . 
\end{eqnarray}

However, the energy-momentum tensor is not symmetric: its antisymmetric part is given by
\begin{eqnarray}
T^{\mu\nu}-T^{\nu\mu}=\lambda\left[u^\mu(\partial^\nu\phi)-u^\nu(\partial^\mu\phi)\right](u\cdot\partial\phi). 
\end{eqnarray}
In principle, this asymmetry is typical for Lorentz-breaking theories.



\section{Lorentz symmetry breaking within the Casimir effect}
\label{chap3}

In this section we will study how the space-time anisotropy generated by the Lorentz-breaking term modifies the results of the Casimir energy associated with a real massive scalar quantum field confined between two large parallel plates. As we have already mentioned, in what follows, we consider the anisotropy  of the space-time generated by a constant four-vector $u^{\mu}$. 
Moreover, we will consider different boundary conditions imposed for the fields $\phi(x)$ on the plates.

\begin{figure}[!htb]
	\label{fig1}
\centering
\includegraphics[scale=0.3]{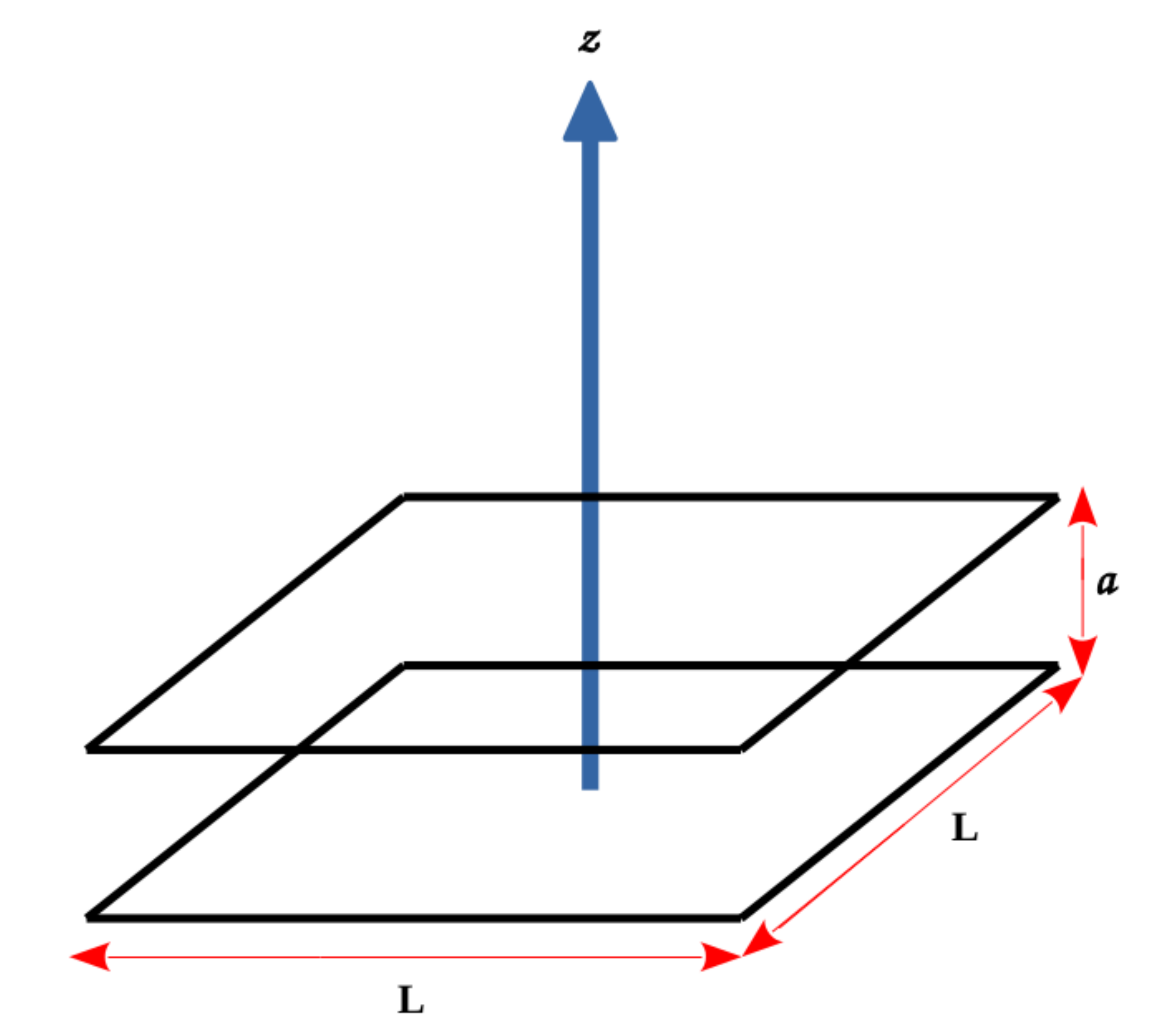}
\caption{Two parallel plates with area $L^{2}$ separated by a distance $a \ll L$.}
\label{placas_paralelas}
\end{figure}

Let us consider a massive scalar quantum field between two large parallel plates, as it is shown in Fig. \ref{fig1}. We saw in the previous section that a real scalar field must satisfy Eq. (\ref{MKG1}).


In our study we will consider the four-vector as being either time-like or space-like. First we must obtain the solution for 
Eq. (\ref{MKG1}) by imposing specific boundary conditions for the fields on the plates and thus obtain the corresponding 
Hamiltonian $\hat{H}$ operator.  Consequently, we can calculate the total vacuum energy of the system and then determine the 
Casimir energy for each of the cases. 

\subsection{Dirichlet boundary condition}
\label{chap3.1}

In this subsection we will solve the modified Klein-Gordon equation \eqref{MKG1} satisfying the Dirichlet boundary conditions given below, 
\begin{eqnarray}
 \left. \phi \left(x\right)\right|_{z=0}=\left. \phi \left(x\right)\right|_{z=a}.
\end{eqnarray}

Adopting the standard procedure \cite{Mandl:2010}, one finds the field operator:
\begin{eqnarray}
 \hat{\phi}(x)=\int{d^{2}\vec{k}}\sum_{n=1}^{\infty}\left[\frac{1}{\left(2\pi\right)^{2}\omega_{\vec{k},n}a}\right]
 ^{\frac{1}{2}}\sin\left(\frac{n\pi}{a}z\right)\left[\hat{a}_{n}(\vec{k})e^{-ikx}+\hat{a}_{n}^{\dagger}(\vec{k})
 e^{ikx}\right],
\end{eqnarray}
where 
\begin{eqnarray}
kx=\omega_{\vec{k},n}t-k_{x}x-k_{y}y.
\end{eqnarray}
With $\hat{a}_{n}(\vec{k})$ and $\hat{a}_{n}^{\dagger}(\vec{k})$ being the annihilation and creation operators,  respectively, characterized by the set of quantum numbers $\sigma=\left\{k_{x},k_{y},n\right\}$. These operators satisfy the algebra

\begin{eqnarray}
\label{vert}
\left \{\begin{array}{c}
 \left[\hat{a}_{n}(\vec{k}),\hat{a}_{n'}^{\dagger}(\vec{k'})\right]=\delta_{n,n'}\delta^{2}(\vec{k}-\vec{k'}), \\ \\
 \left[\hat{a}_{n}(\vec{k}),\hat{a}_{n'}(\vec{k'})\right]=\left[\hat{a}_{n}^{\dagger}(\vec{k}),\hat{a}_{n'}^{\dagger}(\vec{k'})\right]=0. \\
\end{array} \right.
\end{eqnarray}

\subsubsection{Time-like vector case}

Admitting that the four-vector to be time-like, we have $u^{\mu}=(1,0,0,0)$. In this case  we have the dispersion relation given by
\begin{eqnarray}
 \omega^{2}_{\vec{k},n}=\frac{1}{(1+\lambda)}\left[k_{x}^{2}+k_{y}^{2}+\left(\frac{n\pi}{a}\right)^2+m^2\right].
\end{eqnarray}

The Hamiltonian operator, $\hat{H}$, for this case is given by:
\begin{eqnarray}
\begin{aligned}
 \hat{H}=&\frac{1}{2}\int{d^{3}\vec{x}}\left[(1+\lambda)(\partial_{t}\hat{\phi})^{2}+(\vec{\nabla}\hat{\phi})^{2}\right], \\
 =&
 \frac{(1+\lambda)}{2}\int{d^{2}\vec{k}}\sum_{n=1}^{\infty}\omega_{\vec{k},n}\left[2\hat{a}_{n}^{\dagger}(\vec{k})
 \hat{a}_{n}(\vec{k})+\frac{L^2}{(2\pi)^2}\right].
\end{aligned}
\end{eqnarray}

Consequently the vacuum energy is obtained by taking the vacuum expectation value of $\hat{H}$:
\begin{eqnarray}
\label{EVDt}
 E_{0}=\left<\right.0|\hat{H}|\left.0\right>=\frac{(1+\lambda)L^{2}}{8\pi^{2}}\int{d^{2}\vec{k}}\sum_{n=1}^{\infty}
 \omega_{\vec{k},n}.
\end{eqnarray}

In order to develop the summation on the quantum number integer $n$, we shall use the Abel-Plana formula \cite{{Bordag:2009zzd,Saharian:2007ph}}:
\begin{eqnarray}
\label{AP1}
\sum_{n=0}^{\infty}F(n)=\frac{1}{2}F(0)+\int_{0}^{\infty}{dt}F(t)+i\int_{0}^{\infty}\frac{dt}{e^{2\pi t}-1}\left[F(it)-
F(-it)\right].
\end{eqnarray}
Performing in \eqref{EVDt} a change of coordinates in the plate $(k_{x},k_{y})$ to polar ones, we get
\begin{eqnarray}
\label{EC3}
 E_{0}=\frac{(1+\lambda)^{\frac{1}{2}}L^{2}}{4\pi}\int_{0}^{\infty}{kdk}\left[-\frac{1}{2}F(0)+\int_{0}^{\infty}{dt}F(t)+i\int_{0}^{\infty}
 \right.\left.\frac{dt}{e^{2\pi t}-1}\left[F(it)-F(-it)\right]\right]   
\end{eqnarray}
with
\begin{eqnarray}
F(n)=\left[k^2+m^2+\left(\frac{n\pi}{a}\right)^2\right]^{\frac{1}{2}}   \  .
\end{eqnarray}

Note that the first term on the right-hand side of \eqref{EC3} refers to vacuum energy in the presence of only one plate, and the second one is connected with vacuum energy without boundary. Thus, both terms are divergent and do not contribute to Casimir energy. So, we will discard them. As a result, the Casimir energy per unit area of the plates is given by
\begin{eqnarray}
\label{EC2}
 \frac{E_{C}}{L^{2}}=\frac{(1+\lambda)^{\frac{1}{2}}}{4\pi}i\int_{0}^{\infty}{kdk}\int_{0}^{\infty}dt
 \frac{\left[k^2+m^2+\left(\frac{it\pi}{a}\right)^2\right]^{\frac{1}{2}}-\left[k^2+m^2+\left(\frac{-it\pi}{a}\right)^2
 \right]^{\frac{1}{2}}}{e^{2\pi t}-1}.
\end{eqnarray}

Performing a change of variable, where $u=\frac{\pi t}{a}$, we get
\begin{eqnarray}
 \frac{E_{C}}{L^{2}}=\frac{(1+\lambda)^{\frac{1}{2}}a}{4\pi^{2}}i\int_{0}^{\infty}{kdk}\int_{0}^{\infty}du
 \frac{\left[k^2+m^2+(iu)^2\right]^{\frac{1}{2}}-\left[k^2+m^2+(-iu)^2\right]^{\frac{1}{2}}}{e^{2au}-1}.
\end{eqnarray}

The integral over the $u$ variable must be considered in two cases:

\begin{itemize}
 \item For $\left(k^2+m^2\right)^{\frac{1}{2}}>u$:
 \begin{eqnarray}
 \label{ident1}
 \left[k^2+m^2+\left(\pm iu\right)^2\right]^{\frac{1}{2}}=\left[k^2+m^2-u^2\right]^{\frac{1}{2}}.
\end{eqnarray}
\end{itemize}

\begin{itemize}
 \item For $\left(k^2+m^2\right)^{\frac{1}{2}}<u$:
 \begin{eqnarray}
 \label{ident2}
 \left[k^2+m^2+\left(\pm iu\right)^2\right]^{\frac{1}{2}}=e^{\pm i\frac{\pi}{2}}\left[u^2-(k^2+m^2)\right]^{\frac{1}{2}}
\end{eqnarray}
\end{itemize}

Thus integrating can provide two different values. One finds that the integral over $u$ in the interval $[0, (k^2+m^2)^{\frac{1}{2}}]$ 
vanishes. So, we get:
\begin{eqnarray}
 \frac{E_{C}}{L^{2}}=-\frac{(1+\lambda)^{\frac{1}{2}}a}{2\pi^{2}}\int_{0}^{\infty}{kdk}\int_{\sqrt{k^2+m^2}}^{\infty}du
 \frac{\left[u^2-(k^2+m^2)\right]^{\frac{1}{2}}}{e^{2au}-1} \ .
\end{eqnarray}
Performing again another changing of variable, $\rho^{2}=u^{2}-(k^{2}+m^{2})$, we get
\begin{eqnarray}
 \frac{E_{C}}{L^{2}}=-\frac{(1+\lambda)^{\frac{1}{2}}a}{2\pi^{2}}\int_{0}^{\infty}{kdk}\int_{0}^{\infty}{\frac{\rho^{2}d\rho}{
 \sqrt{\rho^{2}+k^{2}+m^{2}}\left(e^{2a\sqrt{\rho^{2}+k^{2}+m^{2}}}-1\right)}} \  .
\end{eqnarray}
Finally doing a change of coordinates in the plate $(k,\rho)$ to polar ones, we get
\begin{eqnarray}
 \frac{E_{C}}{L^{2}}=-\frac{(1+\lambda)^{\frac{1}{2}}a}{6\pi^{2}}\int_{0}^{\infty}\frac{\sigma^{4}d\sigma}{\sqrt{\sigma^{2}+m^{2}}
 \left(e^{2a\sqrt{\sigma^{2}+m^{2}}}-1\right)}.
\end{eqnarray}
Because this integral has no analytic solution, we will consider only asymptotic limit. For this, we will make the following changes of variable $\xi^2=\sigma^2+m^2$ and after that $\xi=mv$, we get
\begin{eqnarray}
\label{int1}
 \frac{E_{C}}{L^{2}}=-\frac{(1+\lambda)^{\frac{1}{2}}am^4}{6\pi^{2}}\int_{1}^{\infty}\frac{(v^{2}-1)^{\frac{3}{2}}dv}
 {e^{2amv}-1}.
\end{eqnarray}

\begin{itemize}
 \item For case $am \gg 1$, we get
 \begin{eqnarray}
 \frac{E_{C}}{L^{2}} \approx -\frac{(1+\lambda)^{\frac{1}{2}}am^{4}}{6\pi^{2}}\int_{1}^{\infty}\frac{(v^{2}-1)^{\frac{3}{2}}dv}
 {e^{2amv}} \approx -\frac{(1+\lambda)^{\frac{1}{2}}}{16}\left(\frac{m}{\pi a}\right)^{\frac{3}{2}}e^{-2ma}.
\end{eqnarray}
\end{itemize}
Notice that in this case the Casimir energy decays exponentially with $ma$.

\begin{itemize}
 \item For case $am\ll 1$, we get
 \begin{eqnarray}
 \begin{aligned}
 \frac{E_{C}}{L^{2}} \approx& -\frac{(1+\lambda)^{\frac{1}{2}}am^{4}}{6\pi^{2}}\int_{1}^{\infty}\frac{v^{3}dv}
 {e^{2amv}-1} \\ 
 \approx& -\frac{(1+\lambda)^{\frac{1}{2}}}{1440\pi^2a^3}\left[\pi^4-40a^3m^3+30a^4m^4-8a^5m^5\right].
 \end{aligned}
\end{eqnarray}
\end{itemize}

It follows from our previous result that the Casimir pressure between two parallel plates arising due to scalar field oscillations takes the form
\begin{eqnarray}
\label{pressure0}
 \begin{aligned}
  P_C(a)=-\frac{(1+\lambda)^{\frac{1}{2}}}{1440\pi^2a^4}\left[3\pi^4-30a^4m^4+16a^5m^5\right].
 \end{aligned}
\end{eqnarray}
So, we conclude that the influence of the Lorentz-symmetry breaking parameter consists only in a multiplicative factor.

\subsubsection{Space-like vector case}

Here we have three different directions for the four-vector $u^{\mu}$, that is: $u^{\mu}=(0,1,0,0)$, $u^{\mu}=(0,0,1,0)$ and 
$u^{\mu}=(0,0,0,1)$. For the two first vectors the dispersion relations are the same. So let us consider $u^{\mu}$ as being,
\begin{eqnarray}
 u^{\mu}=(0,1,0,0).
\end{eqnarray}
The corresponding dispersion relation is:
\begin{eqnarray}
 \omega^{2}_{\vec{k},n}=\left[(1-\lambda)k_{x}^{2}+k_{y}^{2}+\left(\frac{n\pi}{a}\right)^2+m^2\right].
\end{eqnarray}
The Hamiltonian operator, $\hat{H}$ now reads
\begin{eqnarray}
\begin{aligned}
 \hat{H}=&\frac{1}{2}\int{d^{3}\vec{x}}\left[(\partial_{t}\hat{\phi})^{2}+(\vec{\nabla} \hat{\phi})^{2}-\lambda(
 \partial_{x}\hat{\phi})^{2}\right], \\ 
=&\frac{1}{2}\int{d^{2}\vec{k}}\sum_{n=1}^{\infty}\omega_{\vec{k},n}\left[2\hat{a}_{n}^{\dagger}(\vec{k})
 \hat{a}_{n}(\vec{k})+\frac{L^2}{(2\pi)^2}\right].
\end{aligned}
\end{eqnarray}

Consequently the vacuum energy is given by
\begin{eqnarray}
\label{EVDx}
 E_{0}=\left<\right.0|\hat{H}|\left.0\right>=\frac{L^{2}}{8\pi^{2}}\int{d^{2}\vec{k}}\sum_{n=1}^{\infty}
 \omega_{\vec{k},n}.
\end{eqnarray}
In order to sum over $n$ we use again \eqref{AP1}. Proceeding in the same way as before, we obtain
\begin{eqnarray}
\label{int2}
 \frac{E_{C}}{L^{2}}=-\frac{(1-\lambda)^{-\frac{1}{2}}am^{4}}{6\pi^{2}}\int_{1}^{\infty}\frac{(v^{2}-1)^{\frac{3}{2}}dv}
 {e^{2amv}-1}.
\end{eqnarray}

Again considering two asymptotic limits, we have:
\begin{itemize}
 \item For case $am \gg 1$, we get
 \begin{eqnarray}
 \frac{E_{C}}{L^{2}} \approx -\frac{(1-\lambda)^{-\frac{1}{2}}am^{4}}{6\pi^{2}}\int_{1}^{\infty}\frac{(v^{2}-1)^{\frac{3}{2}}dv}
 {e^{2amv}} \approx -\frac{(1-\lambda)^{-\frac{1}{2}}}{16}\left(\frac{m}{\pi a}\right)^{\frac{3}{2}}e^{-2ma}.
\end{eqnarray}
 \item For case $am\ll 1$, we get
 \begin{eqnarray}
 \begin{aligned}
 \frac{E_{C}}{L^{2}} \approx& -\frac{(1-\lambda)^{-\frac{1}{2}}am^{4}}{6\pi^{2}}\int_{1}^{\infty}\frac{v^{3}dv}
 {e^{2amv}-1} \\ 
 \approx& -\frac{(1-\lambda)^{-\frac{1}{2}}}{1440\pi^2a^3}\left[\pi^4-40a^3m^3+30a^4m^4-8a^5m^5\right].
 \end{aligned}
\end{eqnarray}
\end{itemize}

From of the above result, the Casimir pressure between two parallel plates arising due to scalar field oscillations takes the form,
\begin{eqnarray}
\label{pressure1}
 \begin{aligned}
  P_C(a)=-\frac{(1-\lambda)^{-\frac{1}{2}}}{1440\pi^2a^4}\left[3\pi^4-30a^4m^4+16a^5m^5\right].
 \end{aligned}
\end{eqnarray}

Notice that considering only the first order correction in the parameter $\lambda$, the results \eqref{pressure0} and 
\eqref{pressure1} coincide.

Finally, let us consider the four-vector $u^{\mu}$ orthogonal to the plates:
\begin{eqnarray}
 u^{\mu}=(0,0,0,1).
\end{eqnarray}

In this case, the dispersion relation is modified to:
\begin{eqnarray}
 \omega^{2}_{\vec{k},n}=\left[k_{x}^{2}+k_{y}^{2}+(1-\lambda)\left(\frac{n\pi}{a}\right)^2+m^2\right].
\end{eqnarray}
The Hamiltonian operator, $\hat{H}$, now is
\begin{eqnarray}
 \hat{H}&=&\frac{1}{2}\int{d^{3}\vec{x}}\left[(\partial_{t}\hat{\phi})^{2}+(\vec{\nabla}\hat{\phi})^{2}-\lambda(
 \partial_{z}\hat{\phi})^{2}\right], =\nonumber\\
&=&\frac{1}{2}\int{d^{2}\vec{k}}\sum_{n=1}^{\infty}\omega_{\vec{k},n}\left[2\hat{a}_{n}^{\dagger}(\vec{k})
 \hat{a}_{n}(\vec{k})+\frac{L^2}{(2\pi)^2}\right].
\end{eqnarray}
Consequently, one finds that the vacuum energy is given by
\begin{eqnarray}
 E_{0}=\left<\right.0|\hat{H}|\left.0\right>=\frac{L^{2}}{8\pi^{2}}\int{d^{2}\vec{k}}\sum_{n=1}^{\infty}
 \omega_{\vec{k},n}.
\end{eqnarray}
Developing again the summation over the $n$ by using \eqref{AP1} and performing a change of coordinates  $(k_x,k_y)$ to polar coordinates, with
\begin{eqnarray}
 F(n)=\left[k^2+m^2+(1-\lambda)\left(\frac{n\pi}{a}\right)^2\right]^{\frac{1}{2}},
\end{eqnarray}
one finds that the Casimir energy per unit area is given by
\begin{eqnarray}
\label{int3}
 \frac{E_{C}}{L^{2}}=-\frac{bm^{4}}{6\pi^{2}}\int_{1}^{\infty}\frac{(v^{2}-1)^{\frac{3}{2}}dv}
 {e^{2bmv}-1},
\end{eqnarray}
where we identify
\begin{eqnarray}
\label{functionb}
 b=\frac{a}{\left(1-\lambda \right)^{\frac{1}{2}}}.
\end{eqnarray}

The asymptotic limits of the above expression is given below:
\begin{itemize}
 \item For case $am \gg 1$, we get
 \begin{eqnarray}
 \frac{E_{C}}{L^{2}} \approx -\frac{bm^{4}}{6\pi^{2}}\int_{1}^{\infty}\frac{(v^{2}-1)^{\frac{3}{2}}dv}
 {e^{2bmv}} \approx -\frac{(1-\lambda)^{\frac{3}{4}}}{16}\left(\frac{m}{\pi a}\right)^{\frac{3}{2}}e^{-2(1-\lambda)^{-\frac{1}{2}}am}.
\end{eqnarray}
 \item For case $am\ll 1$, we get
 \begin{eqnarray}
 \begin{aligned}
 \frac{E_{C}}{L^{2}} \approx& -\frac{bm^{4}}{6\pi^{2}}\int_{1}^{\infty}\frac{v^{3}dv}
 {e^{2bmv}-1} \\ 
 \approx& -\frac{(1-\lambda)^{-1}}{1440\pi^2a^3}\left[\pi^4(1-\lambda)^{\frac{5}{2}}-40a^3m^3(1-\lambda)+30a^4m^4(1-\lambda)^{\frac{1}{2}}-8a^5m^5 \right].
 \end{aligned}
\end{eqnarray}
\end{itemize}

Therefore, the Casimir pressure between two parallel plates, arising due to scalar field oscillations takes the form,
\begin{eqnarray}
\label{pressure}
 \begin{aligned}
  P_C(a)=-\frac{(1-\lambda)^{-1}}{1440\pi^2a^4}\left[3\pi^4(1-\lambda)^{\frac{5}{2}}-30a^4m^4(1-\lambda)^{\frac{1}{2}}+16a^5m^5\right].
 \end{aligned}
\end{eqnarray}

Here we see that the influence of the Lorentz-violation parameter, $\lambda$, on the Casimir energy is more delicate. 
It appear not only as a multiplicative factor, but also enters the integrand of (\ref{int3}).

\subsection{Neumann boundary condition}

Now we want to obtain solutions of the modified Klein-Gordon equation \eqref{MKG1} which obey the boundary condition below,
\begin{eqnarray}
 \left.\frac{\partial \phi(x)}{\partial z}\right|_{z=0}=\left.\frac{\partial \phi(x)}{\partial z}\right|_{z=a}=0.
\end{eqnarray}

After some intermediate steps, we can say that for this case the field operator reads,
\begin{eqnarray}
 \hat{\phi}(x)=\int{d^{2}\vec{k}}\sum_{n=0}^{\infty}c_{n}\cos \left(\frac{n\pi}{a}z\right)\left[\hat{a}_{n}(\vec{k})
 e^{-ikx}+\hat{a}_{n}^{\dagger}(\vec{k})e^{ikx}\right],
\end{eqnarray}
where the normalization constant is 
\begin{eqnarray}
\label{vert1}
 c_{n}=\left \{\begin{array}{c}
 \left[\frac{1}{2(2\pi)^{\frac{1}{2}}\omega_{\vec{k},n}a}\right]^{\frac{1}{2}} \text{ }\text{for} \text{ }n=0, \\ \\
 \left[\frac{1}{(2\pi)^{\frac{1}{2}}\omega_{\vec{k},n}a}\right]^{\frac{1}{2}} \text{ } \text{for} \text{ } n\geq 0.
\end{array} \right.
\end{eqnarray}

In this case, we notice that the field operator is modified. However, the Hamiltonian operator and the dispersion relations remain the same as to the Dirichlet boundary condition, for each choice of the four-vector $u^{\mu}$, time-like and space-like.

\subsubsection{Time-like vector case}

Considering that the four-vector is time-like, $u^{\mu}=(1,0,0,0)$, we have the following dispersion relation,
\begin{eqnarray}
 \omega^{2}_{\vec{k},n}=\frac{1}{(1+\lambda)}\left[k_{x}^{2}+k_{y}^{2}+\left(\frac{n\pi}{a}\right)^2+m^2\right].
\end{eqnarray}

The Hamiltonian operator, $\hat{H}$, for this case  reads
\begin{eqnarray}
 \hat{H}=\frac{(1+\lambda)}{2}\int{d^{2}\vec{k}}\sum_{n=1}^{\infty}\omega_{\vec{k},n}\left[2\hat{a}_{n}^{\dagger}(\vec{k})
 \hat{a}_{n}(\vec{k})+\frac{L^2}{(2\pi)^2}\right].
\end{eqnarray}
Consequently, the Casimir energy per unit of area is:
\begin{eqnarray}
\label{int11}
 \frac{E_{C}}{L^{2}}=-\frac{(1+\lambda)^{\frac{1}{2}}am^{4}}{6\pi^{2}}\int_{1}^{\infty}\frac{(v^{2}-1)^{\frac{3}{2}}dv}
 {e^{2amv}-1}.
\end{eqnarray}

\begin{itemize}
 \item For case $am \gg 1$, we get
 \begin{eqnarray}
 \frac{E_{C}}{L^{2}} \approx -\frac{(1+\lambda)^{\frac{1}{2}}am^{4}}{6\pi^{2}}\int_{1}^{\infty}\frac{(v^{2}-1)^{\frac{3}{2}}dv}
 {e^{2amv}} \approx -\frac{(1+\lambda)^{\frac{1}{2}}}{16}\left(\frac{m}{\pi a}\right)^{\frac{3}{2}}e^{-2ma}.
\end{eqnarray}
 \item For case $am\ll 1$, we get
 \begin{eqnarray}
 \begin{aligned}
 \frac{E_{C}}{L^{2}} \approx& -\frac{(1+\lambda)^{\frac{1}{2}}am^{4}}{6\pi^{2}}\int_{1}^{\infty}\frac{v^{3}dv}
 {e^{2amv}-1} \\ 
 \approx& -\frac{(1+\lambda)^{\frac{1}{2}}}{1440\pi^2a^3}\left[\pi^4-40a^3m^3+30a^4m^4-8a^5m^5\right].
 \end{aligned}
\end{eqnarray}
\end{itemize}

From of the above result, the Casimir pressure between two parallel plates arising due to scalar field oscillations takes the form,
\begin{eqnarray}
 \begin{aligned}
  P_C(a)=-\frac{(1+\lambda)^{\frac{1}{2}}}{1440\pi^2a^4}\left[3\pi^4-30a^4m^4+16a^5m^5\right].
 \end{aligned}
\end{eqnarray}

\subsubsection{Space-like vector case}

Taking the four-vector as
\begin{eqnarray}
 u^{\mu}=(0,1,0,0)  \  ,
\end{eqnarray}
we obtain the following dispersion relation:
\begin{eqnarray}
 \omega^{2}_{\vec{k},n}=\left[(1-\lambda)k_{x}^{2}+k_{y}^{2}+\left(\frac{n\pi}{a}\right)^2+m^2\right].
\end{eqnarray}
The Hamiltonian operator is given by
\begin{eqnarray}
 \hat{H}=\frac{1}{2}\int{d^{2}\vec{k}}\sum_{n=1}^{\infty}\omega_{\vec{k},n}\left[2\hat{a}_{n}^{\dagger}(\vec{k})
 \hat{a}_{n}(\vec{k})+\frac{L^2}{(2\pi)^2}\right].
\end{eqnarray}
The Casimir energy per unit area is:
\begin{eqnarray}
\label{int22}
 \frac{E_{C}}{L^{2}}=-\frac{(1-\lambda)^{-\frac{1}{2}}am^{4}}{6\pi^{2}}\int_{1}^{\infty}\frac{(v^{2}-1)^{\frac{3}{2}}dv}
 {e^{2amv}-1}.
\end{eqnarray}

\begin{itemize}
 \item For case $am \gg 1$, we get
 \begin{eqnarray}
 \frac{E_{C}}{L^{2}} \approx -\frac{(1-\lambda)^{-\frac{1}{2}}am^{4}}{6\pi^{2}}\int_{1}^{\infty}\frac{(v^{2}-1)^{\frac{3}{2}}dv}
 {e^{2amv}} \approx -\frac{(1-\lambda)^{-\frac{1}{2}}}{16}\left(\frac{m}{\pi a}\right)^{\frac{3}{2}}e^{-2ma}.
\end{eqnarray}
 \item For case $am\ll 1$, we get
 \begin{eqnarray}
 \begin{aligned}
 \frac{E_{C}}{L^{2}} \approx& -\frac{(1-\lambda)^{-\frac{1}{2}}am^{4}}{6\pi^{2}}\int_{1}^{\infty}\frac{v^{3}dv}
 {e^{2amv}-1} \\ 
 \approx& -\frac{(1-\lambda)^{-\frac{1}{2}}}{1440\pi^2a^3}\left[\pi^4-40a^3m^3+30a^4m^4-8a^5m^5\right].
 \end{aligned}
\end{eqnarray}
\end{itemize}

From of the above result, the Casimir pressure between two parallel plates arising due to scalar field oscillations takes the 
form
\begin{eqnarray}
 \begin{aligned}
  P_C(a)=-\frac{(1-\lambda)^{-\frac{1}{2}}}{1440\pi^2a^4}\left[3\pi^4-30a^4m^4+16a^5m^5\right].
 \end{aligned}
\end{eqnarray}
Notice that this result is repeated for the case where $u^{\mu}=(0,0,1,0)$.

Finally, in the case when four-vector $u^{\mu}$ is orthogonal to the plates:
\begin{eqnarray}
 u^{\mu}=(0,0,0,1)   , 
\end{eqnarray}
the dispersion relation is:
\begin{eqnarray}
 \omega^{2}_{\vec{k},n}=\left[k_{x}^{2}+k_{y}^{2}+(1-\lambda)\left(\frac{n\pi}{a}\right)^2+m^2\right].
\end{eqnarray}
The Hamiltonian operator is given by:
\begin{eqnarray}
 \hat{H}=\frac{1}{2}\int{d^{2}\vec{k}}\sum_{n=1}^{\infty}\omega_{\vec{k},n}\left[2\hat{a}_{n}^{\dagger}(\vec{k})
 \hat{a}_{n}(\vec{k})+\frac{L^2}{(2\pi)^2}\right].
\end{eqnarray}
Obtaining the Casimir energy per unit area given by:
\begin{eqnarray}
\label{int33}
 \frac{E_{C}}{L^{2}}=-\frac{bm^{4}}{6\pi^{2}}\int_{1}^{\infty}\frac{(v^{2}-1)^{\frac{3}{2}}dv}
 {e^{2bmv}-1}.
\end{eqnarray}

\begin{itemize}
 \item For case $am \gg 1$, we get
 \begin{eqnarray}
 \frac{E_{C}}{L^{2}} \approx -\frac{bm^{4}}{6\pi^{2}}\int_{1}^{\infty}\frac{(v^{2}-1)^{\frac{3}{2}}dv}
 {e^{2bmv}} \approx -\frac{(1-\lambda)^{\frac{3}{4}}}{16}\left(\frac{m}{\pi a}\right)^{\frac{3}{2}}e^{-2(1-\lambda)^{-\frac{1}{2}}am}.
\end{eqnarray}
 \item For case $am\ll 1$, we get
 \begin{eqnarray}
 \begin{aligned}
 \frac{E_{C}}{L^{2}} \approx& -\frac{bm^{4}}{6\pi^{2}}\int_{1}^{\infty}\frac{v^{3}dv}
 {e^{2bmv}-1} \\ 
 \approx& -\frac{(1-\lambda)^{-1}}{1440\pi^2a^3}\left[\pi^4(1-\lambda)^{\frac{5}{2}}-40a^3m^3(1-\lambda)+30a^4m^4(1-\lambda)^{\frac{1}{2}}-8a^5m^5\right].
 \end{aligned}
\end{eqnarray}
\end{itemize}

From of the above result, the Casimir pressure between two parallel plates due to scalar field oscillations takes the form,
\begin{eqnarray}
 \begin{aligned}
  P_C(a)=-\frac{(1-\lambda)^{-1}}{1440\pi^2a^4}\left[3\pi^4(1-\lambda)^{\frac{5}{2}}
  -30a^4m^4(1-\lambda)^{\frac{1}{2}}+16a^5m^5\right].
 \end{aligned}
\end{eqnarray}

\subsection{Mixed boundary condition}

Now let us consider a scalar field which obeys a Dirichlet boundary condition on one plate, and a Neumann boundary 
condition on the other one. Two different configurations take place:

\begin{itemize}
 \item First configuration,
 \begin{eqnarray}
  \left.\phi(\vec{x})\right|_{z=0}=\left.\frac{\partial \phi(\vec{x})}{\partial z}\right|_{z=a}=0.
 \end{eqnarray}

\end{itemize}

\begin{itemize}
 \item Second configuration,
 \begin{eqnarray}
  \left.\frac{\partial \phi(\vec{x})}{\partial z}\right|_{z=0}=\left.\phi(\vec{x})\right|_{z=a}=0.
 \end{eqnarray}

\end{itemize}

After solving Klein-Gordon Eq. \eqref{MKG1} with these conditions, the field operators read as:
\begin{eqnarray}
\hat{\phi}_{1}(x)=\int{d^{2}\vec{k}}\sum_{n=0}^{\infty}\left[\frac{1}{(2\pi)^{2}\omega_{\vec{k},n}a}\right]^{\frac{1}{2}}
\sin \left[\left(n+\frac{1}{2}\right)\frac{\pi}{a}z\right]\left[\hat{a}_{n}(\vec{k})e^{-ikx}+\hat{a}^{\dagger}_{n}(\vec{k})
e^{ikx} \right],
\end{eqnarray}
for the first configuration and
\begin{eqnarray}
\hat{\phi}_{2}(x)=\int{d^{2}\vec{k}}\sum_{n=0}^{\infty}\left[\frac{1}{(2\pi)^{2}\omega_{\vec{k},n}a}\right]^{\frac{1}{2}}
\cos \left[\left(n+\frac{1}{2}\right)\frac{\pi}{a}z\right]\left[\hat{a}_{n}(\vec{k})e^{-ikx}+\hat{a}^{\dagger}_{n}(\vec{k})
e^{ikx} \right],
\end{eqnarray}
for the second configuration.

Both field operators, $\hat{\phi}_{1}(x)$ and $\hat{\phi}_{2}(x)$, provide the same Hamiltonian operator that satisfies the  same dispersion relations, for the cases of time-like and space-like vectors.

\subsubsection{Time-like vector case}

Suggesting that the four-vector is time-like, $u^{\mu}=(1,0,0,0)$, we have the Hamiltonian operator
\begin{eqnarray}
 \hat{H}=\frac{(1+\lambda)}{2}\int{d^{2}\vec{k}}\sum_{n=0}^{\infty}\omega_{\vec{k},n}\left[2\hat{a}_{n}^{\dagger}(\vec{k})
 \hat{a}_{n}(\vec{k})+\frac{L^{2}}{(2\pi)^{2}}\right],
\end{eqnarray}
where $\omega_{\vec{k},n}$ satisfies the dispersion relation,
\begin{eqnarray}
 \omega^{2}_{\vec{k},n}=\frac{1}{(1+\lambda)}\left[k_{x}^{2}+k_{y}^{2}+\left[\left(n+\frac{1}{2}\right)\frac{\pi}{a}\right]^{2}
 +m^{2}\right].
\end{eqnarray}

The vacuum energy of the scalar field is expressed as
\begin{eqnarray}
 E_{0}=\langle 0|\hat{H}|0 \rangle=\frac{(1+\lambda)L^{2}}{8\pi^{2}}\int{d^{2}\vec{k}}\sum_{n=0}^{\infty}\omega_{\vec{k},n}.
\end{eqnarray}

Changing the coordinates of the plate $(k_{x},k_{y})$ to polar ones, and using the Abel-Plana summation formula for half-integer  numbers \cite{{Bordag:2009zzd,Saharian:2007ph}}:
\begin{eqnarray}
\label{AP2}
 \sum_{n=0}^{\infty}F\left(n+\frac{1}{2}\right)=\int_{0}^{\infty}{F(t)dt}-i\int_{0}^{\infty}{\frac{dt}{e^{2\pi t}+1}\left[F(it)-
 F(-it)\right]} \   ,
\end{eqnarray}
with
\begin{eqnarray}
F\left(n+\frac{1}{2}\right)=\left[k^{2}+m^{2}+\left[\left(n+\frac{1}{2}\right)\frac{\pi}{a}\right]^{2}\right]^{\frac{1}{2}}   \  ,
\end{eqnarray}
one expresses the vacuum energy as
\begin{eqnarray}
\label{ECasimir}
 E_{0}=\frac{(1+\lambda)^{\frac{1}{2}}L^{2}}{4\pi}\int_{0}^{\infty}{kdk}\left[\int_{0}^{\infty}{F(t)dt}-i\int_{0}^{\infty}{dt}
 \frac{F(it)-F(-it)}{e^{2\pi t}+1}\right]  \   .
\end{eqnarray}

Again, the Casimir energy is given by the second term of \eqref{ECasimir}. The first term refers to the free vacuum energy. Then the Casimir energy is given by,
\begin{eqnarray}
 E_{C}=-\frac{(1+\lambda)^{\frac{1}{2}}L^{2}}{4\pi}i\int_{0}^{\infty}{kdk}\int_{0}^{\infty}{dt}\frac{\left[k^{2}+m^{2}+\left(
 \frac{i\pi t}{a}\right)^{2}\right]^{\frac{1}{2}}-\left[k^{2}+m^{2}+\left(-\frac{i\pi t}{a}\right)^{2}\right]^{\frac{1}{2}}}{e^{2\pi t}+1}.
\end{eqnarray}
After performing a change of variable, $\frac{\pi t}{a}=u$, the Casimir energy by unit area reads,
\begin{eqnarray}
 \frac{E_{C}}{L^{2}}=-\frac{(1+\lambda)^{\frac{1}{2}}a}{4\pi^{2}}i\int_{0}^{\infty}{kdk}\int_{0}^{\infty}{dt}\frac{\left[k^{2}+m^{2}+
 \left(iu\right)^{2}\right]^{\frac{1}{2}}-\left[k^{2}+m^{2}+\left(-iu\right)^{2}\right]^{\frac{1}{2}}}{e^{2au}+1}.
\end{eqnarray}

We must consider the integral over the variable $u$ in two sub-intervals: the first one is $[0,(k^2+m^2)^{\frac{1}{2}}]$ and the 
second is $[(k^2+m^2)^{\frac{1}{2}},\infty]$. It follows from \eqref{ident1} that the integral in the interval $[0,(k^2+m^2)^{\frac{1}{2}}]$ vanishes, 
so it remains to study only the integral in the second interval $[(k^2+m^2)^{\frac{1}{2}},\infty]$. By using \eqref{ident2}, we get:
\begin{eqnarray}
 \frac{E_C}{L^2}=\frac{(1+\lambda)^{\frac{1}{2}}a}{2\pi^{2}}\int_{0}^{\infty}{kdk}\int_{\sqrt{k^2+m^2}}^{\infty}{du}
 \frac{\left[u^2-(k^2+m^2)\right]^{\frac{1}{2}}}{e^{2au}+1}.
\end{eqnarray}
Changing the integral coordinate conveniently, we obtain
\begin{eqnarray}
\label{int4}
 \frac{E_C}{L^2}=\frac{(1+\lambda)^{\frac{1}{2}}am^4}{6\pi^{2}}\int_{1}^{\infty}\frac{(v^2-1)^{\frac{3}{2}}}{e^{2amv}+1}{dv}.
\end{eqnarray}
Again, there is no closed expression to the above integral, only asymptotic  expressions can be provided.
\begin{itemize}
 \item For case $am \gg 1$, we get
 \begin{eqnarray}
 \frac{E_{C}}{L^{2}} \approx \frac{(1+\lambda)^{\frac{1}{2}}am^{4}}{6\pi^{2}}\int_{1}^{\infty}\frac{(v^{2}-1)^{\frac{3}{2}}}
 {e^{2amv}}{dv} \approx \frac{(1+\lambda)^{\frac{1}{2}}}{16}\left(\frac{m}{\pi a}\right)^{\frac{3}{2}}e^{-2am}.
\end{eqnarray}
 \item For case $am\ll 1$, we get
 \begin{eqnarray}
 \begin{aligned}
 \frac{E_{C}}{L^{2}} \approx& \frac{(1+\lambda)^{\frac{1}{2}}am^{4}}{6\pi^{2}}\int_{1}^{\infty}\frac{v^{3}}
 {e^{2amv}+1}{dv} \\ \approx& \frac{(1+\lambda)^{\frac{1}{2}}}{11520\pi^2a^3}\left[7\pi^4+48a^4m^4\left(4am-5\right) \right].
 \end{aligned}
\end{eqnarray}
\end{itemize}

From of the above, the Casimir pressure between two parallel plates arising due to scalar field oscillations takes the form
\begin{eqnarray}
 \begin{aligned}
  P_C(a)=\frac{(1+\lambda)^{\frac{1}{2}}}{3840\pi^2a^4}\left[7\pi^4+80a^4m^4-128a^5m^5\right].
 \end{aligned}
\end{eqnarray}
Here also the Lorentz-violation parameter appear only multiplying the standard Casimir pressure.

\subsubsection{Space-like vector case}

Again, we have three different directions for the four-vector $u^{\mu}$. They are: $u^{\mu}=(0,1,0,0)$,  $u^{\mu}=(0,0,1,0)$ 
and $u^{\mu}=(0,0,0,1)$. As we have mentioned the modified Casimir energy associated two  first Lorentz-breaking constant 
vectors provide the same result. So, we will concentrate only with the vector, \begin{eqnarray}
 u^{\mu}=(0,1,0,0)  .
\end{eqnarray}

The Hamiltonian operator, $\hat{H}$, now is
\begin{eqnarray}
 \hat{H}=\frac{1}{2}\int{d^{2}\vec{k}}\sum_{n=0}^{\infty}\omega_{\vec{k},n}\left[2\hat{a}_{n}^{\dagger}(\vec{k})
 \hat{a}_{n}(\vec{k})+\frac{L^2}{(2\pi)^2}\right].
\end{eqnarray}
For this case, the dispersion relation reads,
\begin{eqnarray}
 \omega^{2}_{\vec{k},n}=\left[(1-\lambda)k_{x}^{2}+k_{y}^{2}+\left[\left(n+\frac{1}{2}\right)\frac{\pi}{a}\right]^2+m^2\right] \  .
\end{eqnarray}
Consequently the divergent vacuum energy is given by
\begin{eqnarray}
 E_{0}=\left<\right.0|\hat{H}|\left.0\right>=\frac{L^{2}}{8\pi^{2}}\int{d^{2}\vec{k}}\sum_{n=1}^{\infty}
 \omega_{\vec{k},n}.
\end{eqnarray}

Developing the summation on $n$ by using \eqref{AP2}, and performing a change of coordinates $(k_x,k_y)$ to polar coordinate, where
\begin{eqnarray}
 F\left(n+\frac{1}{2}\right)=\left[k^2+m^2+\left[\left(n+\frac{1}{2}\right)\frac{\pi}{a}\right]^2\right]^{\frac{1}{2}},
\end{eqnarray}
we obtain
\begin{eqnarray}
 E_{0}=\frac{(1-\lambda)^{-\frac{1}{2}}L^{2}}{4\pi}\int_{0}^{\infty}{kdk}\left[\int_{0}^{\infty}{dt}F(t)-i\int_{0}^{\infty}
 \right.\left.\frac{dt}{e^{2\pi t}+1}\left[F(it)-F(-it)\right]\right].
\end{eqnarray}
Proceeding in the same way as before, we obtain for the Casimir energy per unit area,
\begin{eqnarray}
\label{int5}
 \frac{E_{C}}{L^{2}}=\frac{(1-\lambda)^{-\frac{1}{2}}am^{4}}{6\pi^{2}}\int_{1}^{\infty}\frac{(v^{2}-1)^{\frac{3}{2}}dv}
 {e^{2amv}+1}.
\end{eqnarray}

\begin{itemize}
 \item For case $am \gg 1$, we get
 \begin{eqnarray}
 \frac{E_{C}}{L^{2}} \approx \frac{(1-\lambda)^{-\frac{1}{2}}am^{4}}{6\pi^{2}}\int_{1}^{\infty}\frac{(v^{2}-1)^{\frac{3}{2}}dv}
 {e^{2amv}} \approx \frac{(1-\lambda)^{-\frac{1}{2}}}{16}\left(\frac{m}{\pi a}\right)^{\frac{3}{2}}e^{-2ma}.
\end{eqnarray}
 \item For case $am\ll 1$, we get
 \begin{eqnarray}
 \begin{aligned}
 \frac{E_{C}}{L^{2}} \approx& \frac{(1-\lambda)^{-\frac{1}{2}}am^{4}}{6\pi^{2}}\int_{1}^{\infty}\frac{v^{3}dv}
 {e^{2amv}+1} \\ \approx& \frac{(1-\lambda)^{-\frac{1}{2}}}{11520\pi^2a^3}\left[7\pi^4+48a^4m^4\left(4am-5\right) \right].
 \end{aligned}
\end{eqnarray}
\end{itemize}

From of the above, the Casimir pressure between two parallel plates due to scalar field oscillations takes the form,
\begin{eqnarray}
 \begin{aligned}
  P_C(a)=\frac{(1-\lambda)^{-\frac{1}{2}}}{3840\pi^2a^4}\left[7\pi^4+80a^4m^4-128a^5m^5 \right].
 \end{aligned}
\end{eqnarray}

Finally, we consider the four-vector $u^{\mu}$ orthogonal to the plates:
\begin{eqnarray}
 u^{\mu}=(0,0,0,1).
\end{eqnarray}

The Hamiltonian operator, $\hat{H}$, remains the same,
\begin{eqnarray}
 \hat{H}=\frac{1}{2}\int{d^{2}\vec{k}}\sum_{n=1}^{\infty}\omega_{\vec{k},n}\left[2\hat{a}_{n}^{\dagger}(\vec{k})
 \hat{a}_{n}(\vec{k})+\frac{L^2}{(2\pi)^2}\right]  , 
\end{eqnarray}
however, the dispersion relation is modified to:
\begin{eqnarray}
 \omega^{2}_{\vec{k},n}=\left[k_{x}^{2}+k_{y}^{2}+m^2+(1-\lambda)\left[\left(n+\frac{1}{2}\right)\frac{\pi}{a}\right]^2\right].
\end{eqnarray}
Consequently follow that the vacuum energy is given by
\begin{eqnarray}
 E_{0}=\left<\right.0|\hat{H}|\left.0\right>=\frac{L^{2}}{8\pi^{2}}\int{d^{2}\vec{k}}\sum_{n=1}^{\infty}
 \omega_{\vec{k},n}.
\end{eqnarray}

Again repeating the previous steps as before, where now
\begin{eqnarray}
 F\left(n+\frac{1}{2}\right)=\left[k^2+m^2+(1-\lambda)\left[\left(n+\frac{1}{2}\right)\frac{\pi}{a}\right]^2\right]^{\frac{1}{2}},
\end{eqnarray}
we get
\begin{eqnarray}
\label{int6}
 \frac{E_{C}}{L^{2}}=\frac{bm^{4}}{6\pi^{2}}\int_{1}^{\infty}\frac{(v^{2}-1)^{\frac{3}{2}}dv}
 {e^{2bmv}+1},
\end{eqnarray}
with the parameter $b$ given by \eqref{functionb}.
\begin{itemize}
 \item For case $am \gg 1$, we get
 \begin{eqnarray}
 \frac{E_{C}}{L^{2}} \approx \frac{bm^{4}}{6\pi^{2}}\int_{1}^{\infty}\frac{(v^{2}-1)^{\frac{3}{2}}dv}
 {e^{2bmv}} \approx \frac{(1-\lambda)^{\frac{3}{4}}}{16}\left(\frac{m}{\pi a}\right)^{\frac{3}{2}}e^{-2(1-\lambda)^{-\frac{1}{2}}am}.
\end{eqnarray}
 \item For case $am\ll 1$, we get
 \begin{eqnarray}
 \label{energyCb}
 \begin{aligned}
 \frac{E_{C}}{L^{2}} \approx& \frac{bm^{4}}{6\pi^{2}}\int_{1}^{\infty}\frac{v^{3}dv}
 {e^{2bmv}+1} \\ \approx& \frac{(1-\lambda)^{-1}}{11520\pi^2a^3}\left[7\pi^4(1-\lambda)^{\frac{5}{2}}
 -240a^4m^4(1-\lambda)^{\frac{1}{2}}+192a^5m^5\right].
 \end{aligned}
\end{eqnarray}
\end{itemize}

From of the above, the Casimir pressure between two parallel plates due to scalar field oscillations takes the form,
\begin{eqnarray}
\label{pressure2}
 \begin{aligned}
  P_C(a)=\frac{(1-\lambda)^{-1}}{3840\pi^2a^4}\left[7\pi^4(1-\lambda)^{\frac{5}{2}}+80a^4m^4(1-\lambda)^{\frac{1}{2}}-128a^5m^5\right].
 \end{aligned}
\end{eqnarray}
As in  \eqref{pressure}, the modification in the Casimir energy arising due to the parameter $\lambda$ turns out to be more 
profound than a multiplicative factor.

As we can see the results obtained in this subsection differ from the results found for the Dirichlet and Neumann boundary conditions, 
by a numerical factor and by the opposite sign. So the Casimir energy strongly depends on the boundary conditions imposed to the field.

To illustrate our conclusion, here we present the numerical solutions of the integrals as function of $am$ considering 
different values of the parameter $\lambda$. As we have already mentioned, unfortunately there are no analytical expressions 
for the integrals \eqref{int1}, \eqref{int2} and \eqref{int3} related with the both Dirichlet and Neumann boundary 
conditions, as well for the integrals \eqref{int4}, \eqref{int5} and \eqref{int6} associated with the mixed condition. The 
only possible way is to provide asymptotic expression in the limit of small and large $am$.

\newpage
\begin{itemize}
 \item Dirichlet and Neumann boundary conditions
\end{itemize}
\begin{figure}[!htb]
\centering
\includegraphics[scale=0.7]{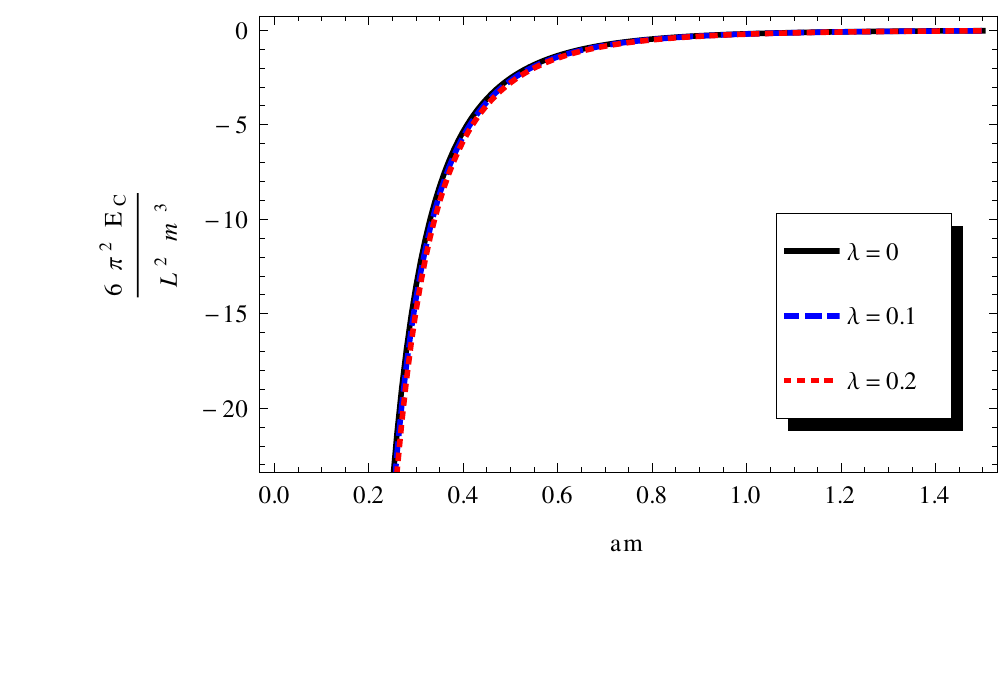}
\caption{Casimir energy in case $u^{\mu}=(1,0,0,0)$ multiplied conveniently by $\frac{6\pi^{2}}{L^2m^3}$, as a function of 
$am$ for the values of parameter $\lambda=0.0, \ \  0,1, \ \ 0.2$.}
\label{Graf_Drich_Neum_tt}
\end{figure}
\begin{figure}[!htb]
\centering
\includegraphics[scale=0.7]{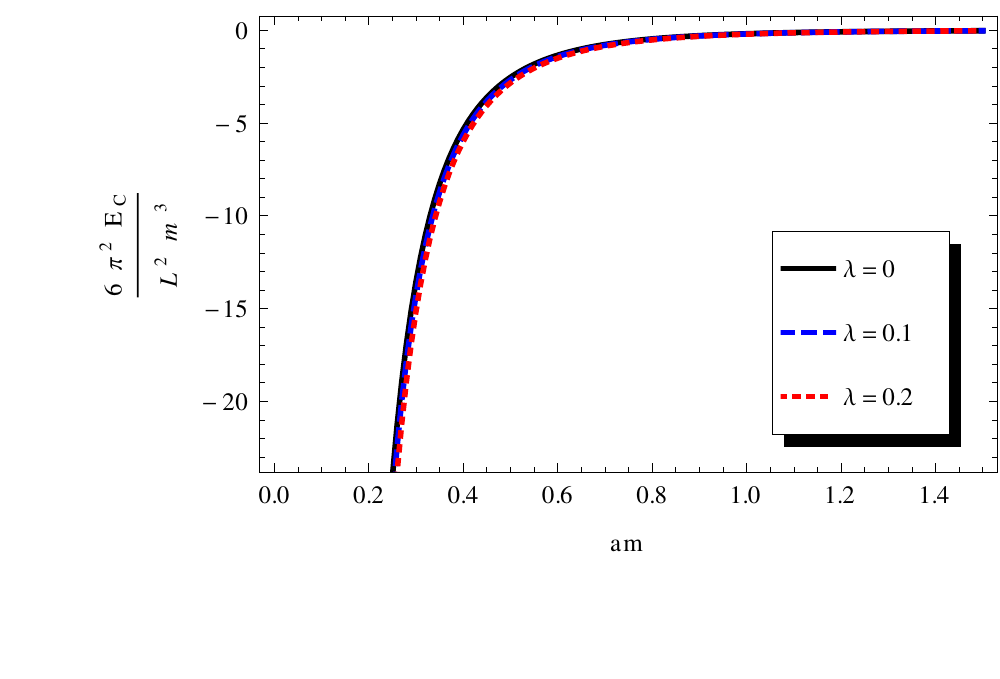}
\caption{Casimir energy in case $u^{\mu}=(0,1,0,0)$ multiplied conveniently by $\frac{6\pi^{2}}{L^2m^3}$, as a function of 
$am$ for the values of parameter $\lambda$.}
\label{Graf_Drich_Neum_texy}
\end{figure}
\begin{figure}[!htb]
\centering
\includegraphics[scale=0.7]{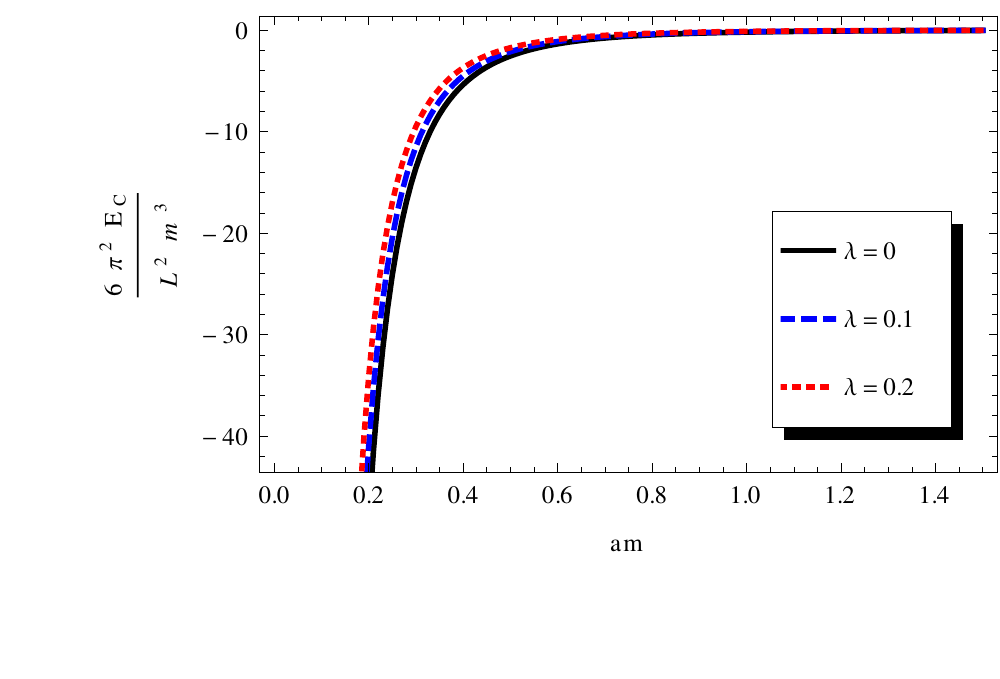}
\caption{Casimir energy in case $u^{\mu}=(0,0,0,1)$ multiplied conveniently by $\frac{6\pi^{2}}{L^2m^3}$, as a function of 
$am$ for the values of parameter $\lambda$.}
\label{Graf_Drich_Neum_tez}
\end{figure}

\newpage
\begin{itemize}
 \item Mixed boundary condition
\end{itemize}
\begin{figure}[!htb]
\centering
\includegraphics[scale=0.7]{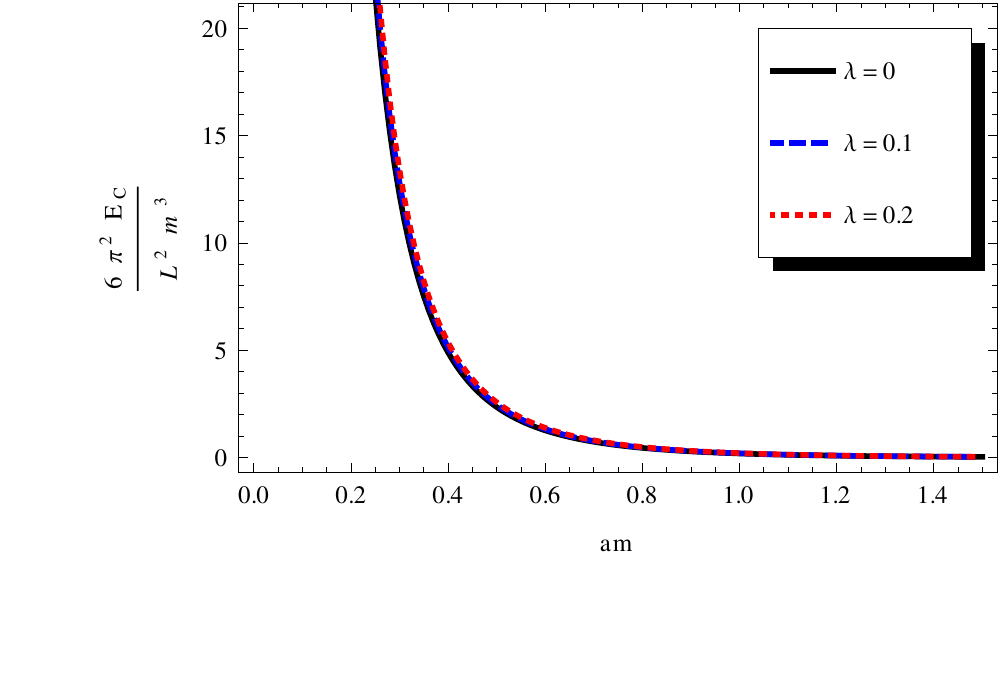}
\caption{Casimir energy in case $u^{\mu}=(1,0,0,0)$ multiplied conveniently by $\frac{6\pi^{2}}{L^2m^3}$, as a function of 
$am$ for the values of parameter $\lambda$.}
\label{Graf_Mista_tt}
\end{figure}
\begin{figure}[!htb]
\centering
\includegraphics[scale=0.7]{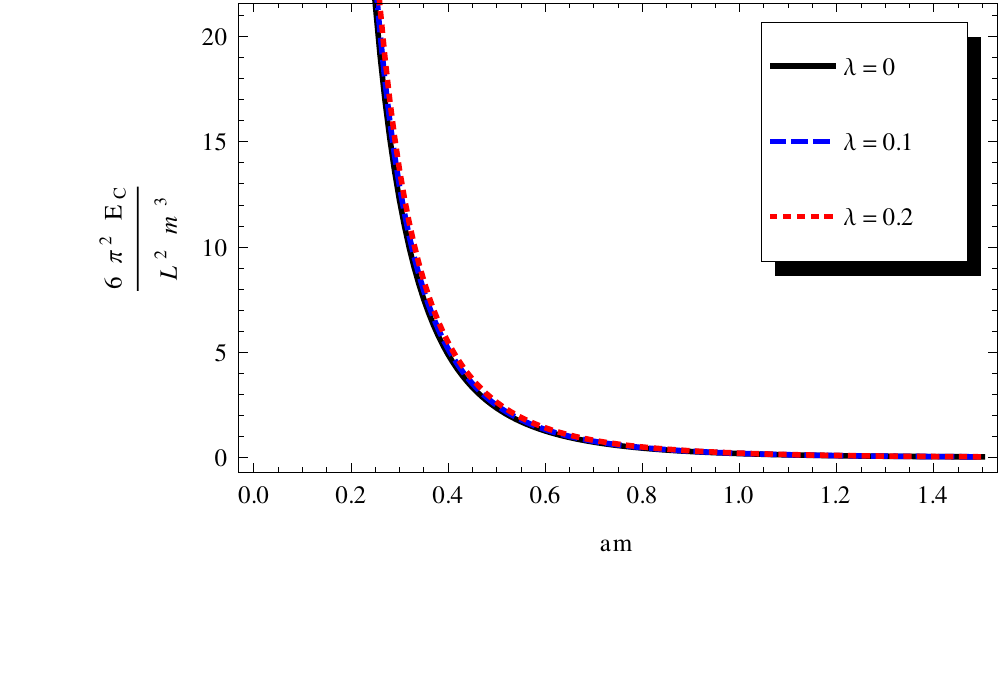}
\caption{Casimir energy in case $u^{\mu}=(0,1,0,0)$ multiplied conveniently by $\frac{6\pi^{2}}{L^2m^3}$, as a function of 
$am$ for the values of parameter $\lambda$.}
\label{Graf_Mista_texy}
\end{figure}
\begin{figure}[!htb]
\centering
\includegraphics[scale=0.7]{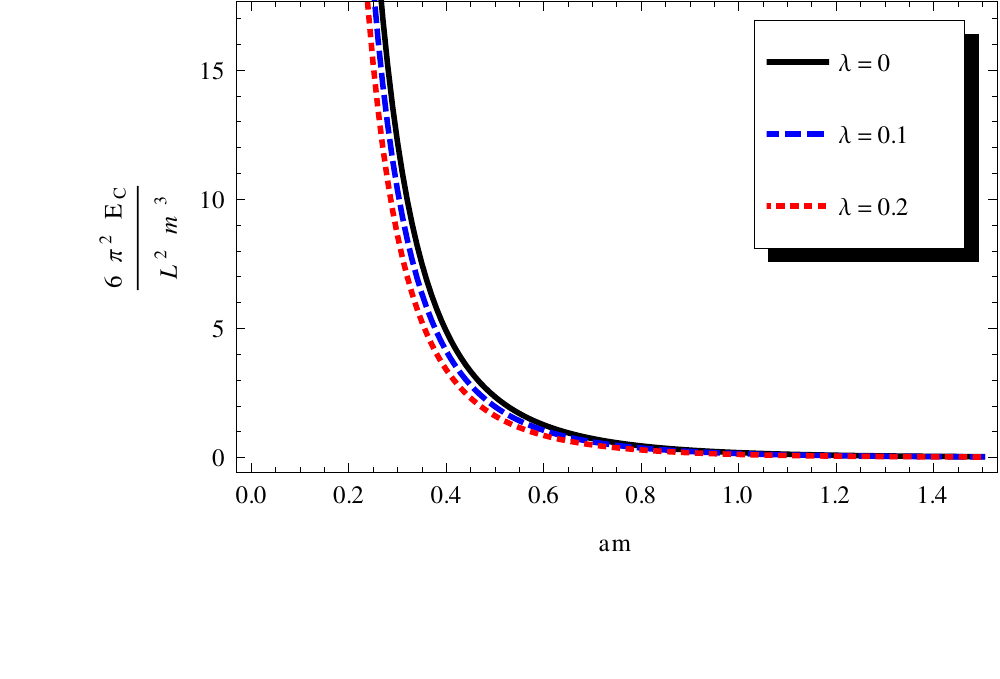}
\caption{Casimir energy in case $u^{\mu}=(0,0,0,1)$ multiplied conveniently by $\frac{6\pi^{2}}{L^2m^3}$, as a function of 
$am$ for the values of parameter $\lambda$.}
\label{Graf_Mista_tez}
\end{figure}

\newpage
\section{Concluding Remarks}

In this work we have investigated the Casimir effect associated to a real massive scalar quantum field in a theoretical model considering an aether-like CPT-even Lorentz symmetry breaking by admitting a direct coupling between the derivative of the  field and an arbitrary constant vector. We considered the situation in which the field is confined between two parallel plates and it was assumed  that the field obeys boundary conditions of the type: Dirichlet, Neumann or mixed ones, with each plate has an area $L^2$, 
and the distance between the plates is $a\text{ }(a\ll L)$.

For each arbitrary direction of the constant vector, $u^{\mu}$, the Casimir energy was obtained. It was observed that in all cases, it is given by a divergent sum, but 
using the Abel-Plana sum formulas Eqs. \eqref{AP1} and \eqref{AP2} allow to obtain finite quantity. 

Considering the Dirichlet and Neumann boundary conditions, it was observed that this sum has three contributions: the contribution from free vacuum energy 
(without plates), the contribution from vacuum energy in the presence of only one plate and the contribution of energy of the  vacuum in the presence of two plates, named the Casimir energy.  The Casimir energies for the Dirichlet and Neumann boundary conditions are given by Eqs. \eqref{int1}, \eqref{int2},  \eqref{int3}, \eqref{int11}, \eqref{int22} and \eqref{int33} for different choices of the four-vector $u^{\mu}$ 
(time-like and space-like). 

For the mixed boundary conditions, it was observed that the sum has only two contributions: the contribution from vacuum energy in the presence of one plate and the contribution from vacuum energy in the presence of  two plates. However, the contributions of free vacuum energy and the contribution of energy in the presence of just one plate  are infinite terms that are subtracted by the renormalization process, resulting in Casimir energy,  given by Eqs. \eqref{int4}, \eqref{int5} and \eqref{int6} for different choices of the four-vector $u^{\mu}$  (time-like and space-like).

As in the usual case, where the Lorentz symmetry is preserved, the Casimir energy in the Dirichlet and Neumann conditions are equal and differs from the Casimir energy in the mixed conditions by a numerical factor and also by a change of the signal.

In all cases it was observed that the Casimir energy strongly depends on the parameter $\lambda$, and for $\lambda=0$ the results are recovered where the Lorentz symmetry is preserved. Therefore, the Casimir energy depends both on the boundary 
conditions imposed on the fields, as well as on the Lorentz-breaking parameter.

In \cite{FrankTuran}, the Casimir effect has been studied for the CPT-even Lorentz-breaking extension of the Maxwell Lagrangian. There, the authors claimed that the dynamics for each of two physical degrees of freedom of the electromagnetic field, is similar to the dynamics of the massless scalar field. In this sense one may infer about some similarities between their results and ours. In order to clarify this point, here we want to point out that the main differences between our results from those ones are based on following reasons. First, we considered the massive scalar field, while the electromagnetic field, in the CPT-even case, is essentially massless; its action does not involve any constants with a nontrivial mass dimension, while the result for the Casimir force in our case is a series in a product $ma$, for the small values of this parameter, and only the lowest order in this expansion, that one corresponding to zero, displays a certain similarity with the result obtained in \cite{FrankTuran}, that is, difference with the Lorentz-invariant case by the constant multiplicative factor which in our case looks like $(1\pm\lambda)^n$ with $|\lambda|\ll 1$ measuring the intensity of the Lorentz symmetry breaking, with different $n$ for different boundary conditions and directions of the constant vector. Second, we considered not only the Dirichlet boundary conditions on the scalar fields as in \cite{FrankTuran}, but also the Neumann and mixed ones. Finally, we note that actually the parameters $\rho$ and $\sigma$ describing propagation of waves in \cite{FrankTuran} are only approximately constants since they obtained from contractions of the object $\kappa^{\alpha\mu\beta\nu}\frac{p_{\mu}p_{\nu}}{\vec{p}^2}$, with $\kappa^{\alpha\mu\beta\nu}$ constant, thus depending on momenta, while our results are perfect constants, being functions only of the dimensionless $\lambda$ without any dependence on momenta.    
To close the paper, we note that, in principle, if the detailed observations and measurements for the scalar field could be possible, one could use the modifications of the Casimir effect  to estimate values of the parameter describing the space-time anisotropy thus contributing to experimental measuring of the Lorentz-breaking 
parameters. 

{\bf Acknowledgements.} This work was partially supported by Conselho
Nacional de Desenvolvimento Cient\'{\i}fico e Tecnol\'{o}gico (CNPq). A. Yu. P. has been partially supported by the CNPq 
through the project No. 303783/2015-0, E. R. Bezerra de Mello through the project 
No. 313137/2014-5. M. B. Cruz has been supported by  
Coordenação de Aperfeiçoamento de Pessoal de Nível Superior (CAPES).


\end{document}